\documentstyle[12pt,aaspp4]{article}
\begin{document}

\title{Revisiting the Modified Eddington Limit for Massive Stars}
\author{Andrew Ulmer\altaffilmark{1} and
Edward L. Fitzpatrick\altaffilmark{2}}
\affil{Princeton University Observatory, Peyton Hall, Princeton, NJ 08544}

\altaffiltext{1}{andrew@astro.princeton.edu}
\altaffiltext{2}{fitz@astro.princeton.edu, current address:
Villanova University, 800 Lancaster Ave. Villanova, PA 19085}
 
\centerline{submitted to {\it ApJ}, August 1997}

\begin{abstract}

We have determined the location of the line-opacity modified Eddington
limit for stars in the LMC using the most recent atmosphere models
combined with a precise mapping to the HR Diagram through up-to-date
stellar evolution calculations.
While we find, in agreement with previous studies,
that the shape of the modified Eddington limit {\em qualitatively} corresponds
to the Humphreys-Davidson (HD) limit defined by the most luminous
supergiants, the modified limit is actually {\it a full magnitude higher}
than the upper luminosity limit observed for LMC stars.
The observed limit is
consistent with atmosphere models in which the maximum value of the
ratio of the radiation force outwards to the gravitational force
inwards, $Y_{\rm max}$, is 0.9, i.e., the photospheres of stars at
the observed luminosity limit are bound.  
As massive stars evolve, they move to higher, and therefore less stable
values of $Y_{\rm max}$, so mass loss, either sporadic or continuous, may
halt their natural redward evolution as they approach the
$Y_{\rm max} = 0.9$ limit.
We assess the metallicity dependence of this limit.
If the limit does determine the most luminous stars, and
the value of $Y_{\rm max}$ corresponding to
the luminosity limit in the LMC is universal, then
the brightest supergiants the SMC should be only marginally brighter
(0.3~mag) than those of the LMC, in agreement with observations.
Moreover, the brightest supergiants in M31 should be 0.75~mag fainter
than those in the LMC.

\end{abstract}


\section{Introduction}

The existence of a temperature-dependent upper-luminosity limit for
massive stars was first pointed out by Hutchings (1976) from
observations of the Large Magellanic Cloud (LMC), and subsequently
quantified by Humphreys \& Davidson (1979).  The essence of this limit,
often referred to as the ``HD-limit,'' is that the maximum luminosity
observed for O-type stars is considerably higher than that seen for the
most luminous M-supergiants.  This observation has had a profound
impact on our understanding of the evolution of massive stars,
indicating that stars with initial masses greater than $\sim$40
$M_{\odot}$ spend their lives in the blue part of the HR Diagram without
becoming red supergiants.

Evolution models for massive stars can be made to reproduce the
observed HR Diagram if sufficiently high stellar mass loss rates are
assumed near the HD-limit (e.g., see Schaller~et~al.~1992).  High mass
loss leads to the removal of the H-rich stellar envelopes, halts
redward evolution, and ultimately produces Wolf-Rayet stars.  The
presence near the observed HD-limit of the luminous blue variable stars
(LBVs), with their occasional outbursts and extreme mass ejections,
seems to verify that mass loss --- perhaps violent and episodic --- is
indeed the primary agent for shaping the properties of the upper HR
Diagram (see Humphreys \& Davidson 1994 for a detailed review of the
LBVs).

One of the most fundamental unanswered questions regarding the
evolution of massive stars is the nature of the underlying
instability mechanism which induces mass loss rates high enough to
produce the outbursts observed in the LBVs and to carve the HD-limit
into the HR Diagram.  One of the first suggestions was that stars
become unstable near the HD-limit due to radiation pressure (Humphreys
\& Davidson 1984; Lamers 1986; Appenzeller 1986).  This model holds
that as massive stars ($M > 40 M_{\odot}$) evolve away from the Zero
Age Main Sequence their photospheres become decreasingly stable against
radiation pressure and ultimately reach a critical point where the
radiation pressure and gravity are balanced, leading to large mass loss
and ending the redward evolution.  Because of metal line opacity,
the luminosity at which a stellar photosphere 
becomes unstable is much lower than that predicted by the classical
electron-scattering Eddington limit.
 
Quantitative studies of a ``modified Eddington limit'' were performed
by Lamers \& Fitzpatrick (1988, hereafter LF) using low gravity,
line-blanketed, plane-parallel, LTE model atmosphere calculations.  By
extrapolating from the low gravity models to a point at which radiation
pressure balanced gravitational pressure, they determined that the
modified Eddington limit was in reasonable agreement with the observed
upper luminosity limit for hot stars ($>10,000$~K) in the LMC.  Lamers
\& Noordhoek (1993, hereafter LN) extended this work to examine the
metallicity dependence of the modified Eddington limit; Achmad,
de~Jager, and Nieuwenhuijzen (1993) found that cool supergiants
($<10,000$~K) are observationally excluded from the region of
luminosity/temperature space predicted to be unstable from the modified
Eddington limit approach (\cite{gus92}).

Alternate explanations for the HD-limit include instabilities of radial
modes in massive stars (\cite{gla93}; \cite{kir93}); turbulent pressure
(e.g., de Jager 1984); and binary star models
(e.g., Kenyon \& Gallagher 1985).  Humphreys \& Davidson
(1994) critically review all these proposed instability mechanisms and
conclude that none, at least in the current state of development, is
fully satisfactory.  It is important to understand the nature of the
mass loss and instability mechanisms operating in the upper HR diagram
--- not only to complete the theoretical picture of stellar evolution,
but also to aid in the interpretation of observations of massive
stars.  Perhaps the most obvious application of such an understanding
would be to determine whether the brightest stars can be used as
reliable distance indicators (e.g., \cite{hum87}).

In this paper, we revisit the modified Eddington limit scenario
proposed by Lamers and collaborators.  We utilize up-to-date stellar
atmosphere calculations to evaluate the radiation pressure stability of
low surface gravity stars and the most recent stellar evolution
calculations to transform the stellar atmosphere parameters
($T_{\rm eff}$ and $\log g$) to the HR diagram ($T_{\rm eff}$ and $L$).
The model atmosphere calculations and the transformation to the HR Diagram
are described in \S~\ref{lowgsec}. In \S~\ref{comobssec},
the modified Eddington limit is compared with the observed upper HR
Diagram of the LMC. Concluding remarks are given in \S~\ref{concomsec}.

\section{Low Gravity Stellar Photosphere Models and the HR Diagram
\label{lowgsec}}

As in LF, our basic procedure is to compute line-blanketed,
plane-parallel, LTE stellar photosphere models for many $T_{\rm eff}$'s
corresponding to OB stars. We compute the models at each temperature,
for surface gravities ranging from those appropriate for the main
sequence ($\log g$ $\simeq$ 4.0) down to the lowest values for
which a model in hydrostatic equilibrium can be computed.
We then determine the
luminosity, $L$, corresponding to each model atmosphere from stellar
evolution calculations and thus can place the atmosphere models on the
HR Diagram ($L$ vs. $T_{\rm eff}$) and compare with observations.

For calculating stellar atmospheres, we employ the ATLAS9 model
atmosphere code of Kurucz (1995), kindly provided by R.L. Kurucz.  The
opacity distribution functions (ODF's) needed to compute the models
were obtained from the CCP7 library (Jeffery 1990).  We produced grids
of models for four different metallicities, $Z/Z_{\odot}$ = 2.0, 1.0,
0.3, and 0.1 (appropriate for M31, the Milky Way, LMC, and SMC) using
the ODF's corresponding to a microturbulence velocity of $v_{\rm turb}
= 8$ km s$^{-1}$.  In a study of the energy distributions of early-type
stars using low-dispersion $IUE$ data, Fitzpatrick \& Massa (in
progress) find that such large values of $v_{\rm turb}$ are required to
reproduce the observed UV opacity in O stars and high-luminosity B
stars.
It is likely that the large equivalent widths of the
strong stellar ``photospheric'' absorption features, which
require the high values of $v_{\rm turb}$ to reproduce, are actually
caused by a physical mechanism very different from microturbulence,
namely, systematic velocity gradients due to increasingly deep
penetration of the stellar wind into the photosphere (e.g., Massa,
Shore, \& Wynne 1992). Nevertheless, the important point for this
investigation is that the $v_{\rm turb} = 8$ km s$^{-1}$ models
represent the {\it observed opacities} remarkably well.

In all, we computed several thousand low gravity models
(which are available on request) at 35 different values
of $T_{\rm eff}$ between 10,000 K and
50,000 K.  We characterize each atmosphere with the parameter $Y_{\rm
max}$ --- suggested by Humphreys \& Davidson (1994) --- which is the
maximum value of the ratio of the outward radiative force to the inward
Newtonian gravitational force found within the optical depth range
$10^{-2} < \tau < 10^2$, i.e.,
$Y_{\rm max} = g_{\rm rad, max}/g_{\rm grav}$.
A value of $Y_{\rm max} = 1$, which defines the modified Eddington limit,
corresponds to the case where the radiative and gravitational forces
are equal.
  
A model in hydrostatic equilibrium cannot be computed for
$Y_{\rm max} = 1$, nor can models be found arbitrarily close to
this value.
For most models, $Y_{\rm max}$ is obtained at optical depths of $\tau
\simeq$ 1--15.  However, close to the modified
Eddington limit the region with highest radiative acceleration
generally shifts to the surface, at
$\tau < 10^{-3}$, and this constrains the lowest surface
gravity for which a hydrostatic model can be computed.
As noted by LF
and LN, this is not considered to represent the modified Eddington
limit because the densities at these surface points are so low that even
an atmosphere which is not in hydrostatic equilibrium at the surface
would add essentially nothing to the mass loss and
because the tops of the photospheres of normal ``stable'' OB stars
merge with the stellar winds and are not in hydrostatic equilibrium.
Both LF and LN extrapolated
to estimate the value of $g$ corresponding
to the hypothetical case where $Y_{\rm max} = 1$ (see Fig. 3 in LF)
from models, in which $Y_{\rm max}$ was determined at $\tau > 10^{-2}$.
In this paper, we restrict our attention to values of $Y_{\rm max}$
less than 0.95; extrapolations are required only in a small number cases
and will be noted where appropriate.

\begin{figure}
\plotone{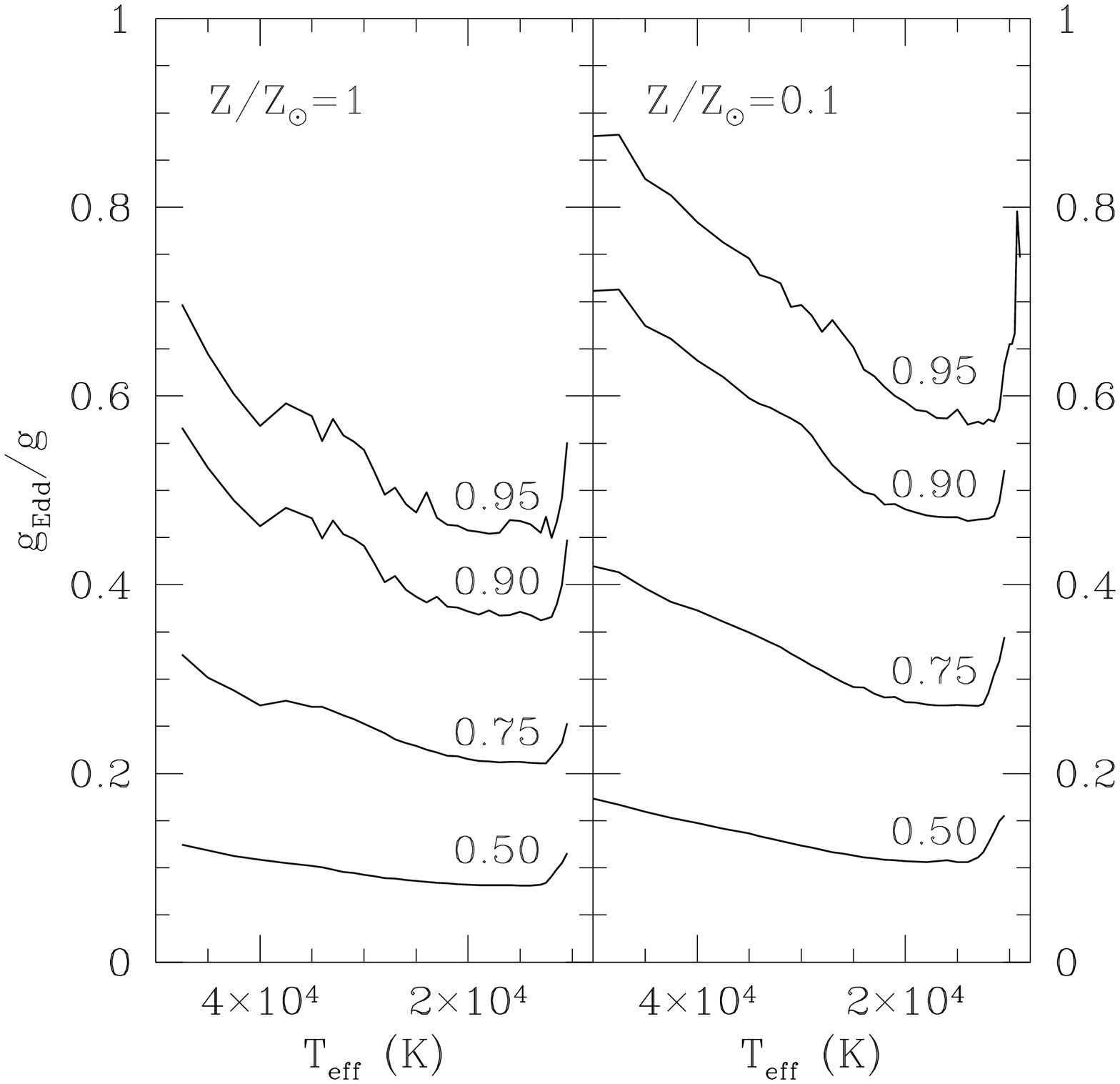}
\caption{\label{ggedd}
The gravities derived from our grid of log-gravity models for
which $Y_{\rm max} =$ 0.95, 0.90, 0.75, and 0.50,
where $Y_{\rm max}$ is the maximum value of the ratio of the outward
radiative force to the inward Newtonian gravitational force found
within the optical depth range $2\times10^{-2} < \tau < 10^2$, i.e.
$Y_{\rm max} = g_{\rm rad, max}/g_{\rm grav}$.
The value of $Y_{\rm max} = 0.90$
most closely follows the HD-limit for the LMC.
The limiting luminosities are significantly
lower than those of the electron-scattering Eddington limit
and correspond to $0.3 - 0.5 L_{\rm Edd}$ and $2-3 g_{\rm Edd}$,
where $g_{\rm Edd} = {4 \pi \sigma T_{\rm eff}^4 G /(L_{\rm Edd}/M_\star)}$.}
\end{figure}

Figure~\ref{ggedd} demonstrates how the surface gravity, $g$, of models
approaching the modified Eddington limit compare to those at the
classical electron-scattering Eddington limit, defined by
\begin{equation} 
g_{\rm Edd} = {4 \pi \sigma T_{\rm eff}^4 G \over
L_{\rm Edd}/M_\star} \approx 6.55 \times 10^{-16} T_{\rm eff}^4
\left({\mu_e \over 1.15}\right)^{-1}~ {\rm cm~s^{-2}}, 
\end{equation}
where $\mu_{\rm e}$ is the mean atomic weight per electron.  The two
panels show the ratio of the surface gravities over a range of
effective temperatures for two different metallicities and four
representative values of $Y_{\rm max}$ (0.95, 0.90, 0.75, and 0.50).
At all metallicities, the highest temperature
models come closest to $g_{\rm Edd}$ since their atmospheric
opacity is dominated by electron scattering.  At lower temperatures,
metal line-blanketing in the UV becomes increasingly important and the
classical and modified limits diverge.  The rise in $g_{\rm Edd}/g$
below about 11,000 K is likely caused by the shifting of the emergent
energy distributions out of the UV and into the relatively unblanketed
optical region.  As expected, the modified Eddington limit comes
closest to the electron scattering limit in the lowest metallicity
models. Modest extrapolation is required outside the range,
$\sim 13,000-30,000$K to reach $Y_{\rm max}=0.95$.
The extrapolation is largest for models with $T_{\rm eff} > 40,000$~K and
solar metallicity.

To determine the luminosities corresponding to our atmosphere models,
we use the stellar evolution grids published by Schaller et al. (1992,
for $Z/Z_{\odot}$ = 1.0 and 0.05), Schaerer et al. (1993a, for
$Z/Z_{\odot}$ = 0.4), Schaerer et al. (1993b, for $Z/Z_{\odot}$ = 2.0),
and Charbonnel et al. (1993; for $Z/Z_{\odot}$ = 0.2).  These models
were computed with the most recent updates of the relevant physical
parameters (e.g., opacities), include the effect of mass loss by winds,
and were tabulated explicitly for ease of interpolation within and
between the grids.  For simplicity, both LF and LN used a
mass-luminosity relation based on the end of the core hydrogen burning
evolutionary phase (CHB) to map the atmosphere models onto the HR
Diagram.  LN noted that many of the models considered actually
corresponded to stars still in the CHB phase, and that this procedure
limits the ability to make quantitative comparisons with observations.
We take a different approach here and interpolate within a grid of
stellar evolution calculations (of the appropriate metallicity) to find
the initial masses of all models which pass through a given set of
$T_{\rm eff}$ and $\log g$ values, as well as the stellar luminosity at
the desired $T_{\rm eff}$ and $\log g$.  In this way we achieve an
essentially exact mapping of the $T_{\rm eff}$ and $\log g$ values onto
the HR Diagram without simplifying assumptions.

\begin{figure}
\plotfiddle{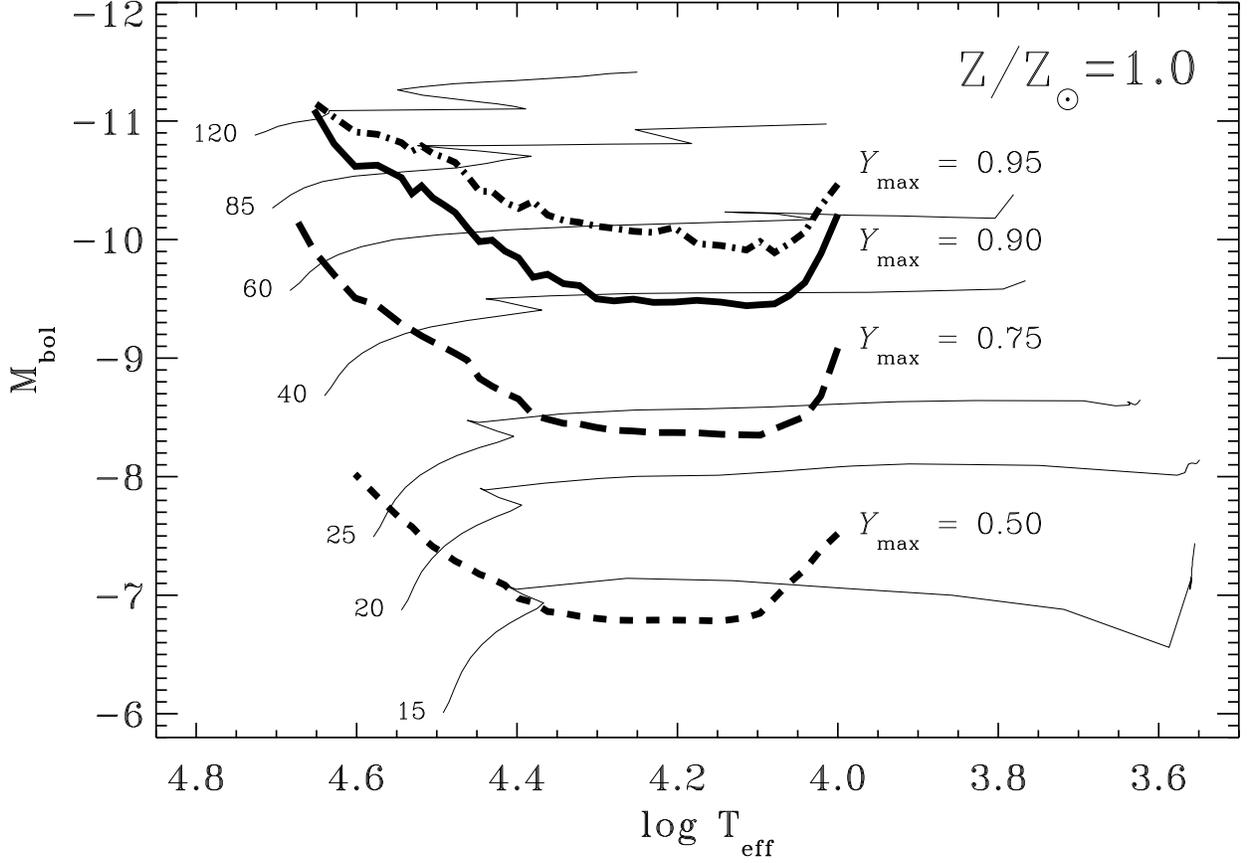}{5.0 in}{90.}{88.0}{90.0}{352.}{-80.}
\caption{\label{soltracks}
The locus of atmosphere models for representative values of
$Y_{\rm max} = g_{\rm rad}/g_{\rm grav}$ = 0.95, 0.90,
0.75, and 0.50, are shown as thick solid and dashed lines.
The thin lines show evolution tracks and
are labeled with their respective initial masses (in units of
$M_{\odot}$) (\cite{sch92}).
For the models with $M_{\rm i} < 25 M_{\odot}$ we show the
tracks from the Zero Age Main Sequence to the end of core helium
burning.  For the more massive stars we truncate the tracks at the
coolest point in the evolution, before the tracks double back to the
blue.}
\end{figure}

Figure~\ref{soltracks} shows the results of this mapping onto the
$M_{bol}$ vs. $\log T_{eff}$ diagram for the calculations done with
solar metallicity for representative values of $Y_{\rm max} = g_{\rm
rad, max}/g_{\rm grav}$ = 0.95, 0.90, 0.75, and 0.50.
Figure~\ref{soltracks} yields two important results.
First, the shape of the curves with $Y_{\rm max}$ values less than 1
are very similar to each other and to that derived by LN for the
extrapolated case of $Y_{\rm max}$ = 1.0.  This characteristic shape,
dubbed the ``Eddington trough'' by LN, is thus not unique to the
hypothetical point of radiative instability, but rather represents the
locus of constant $Y_{\rm max}$ values.  Second, during the CHB phase
the atmospheres of massive stars evolve in the direction of increasing
$Y_{\rm max}$, i.e., towards decreased stability against radiation
pressure.  These points will be discussed further in the following
section, where the modified Eddington limit is compared with
observations.

\section{Comparison With Observations \label{comobssec}}

\begin{figure}
\plotfiddle{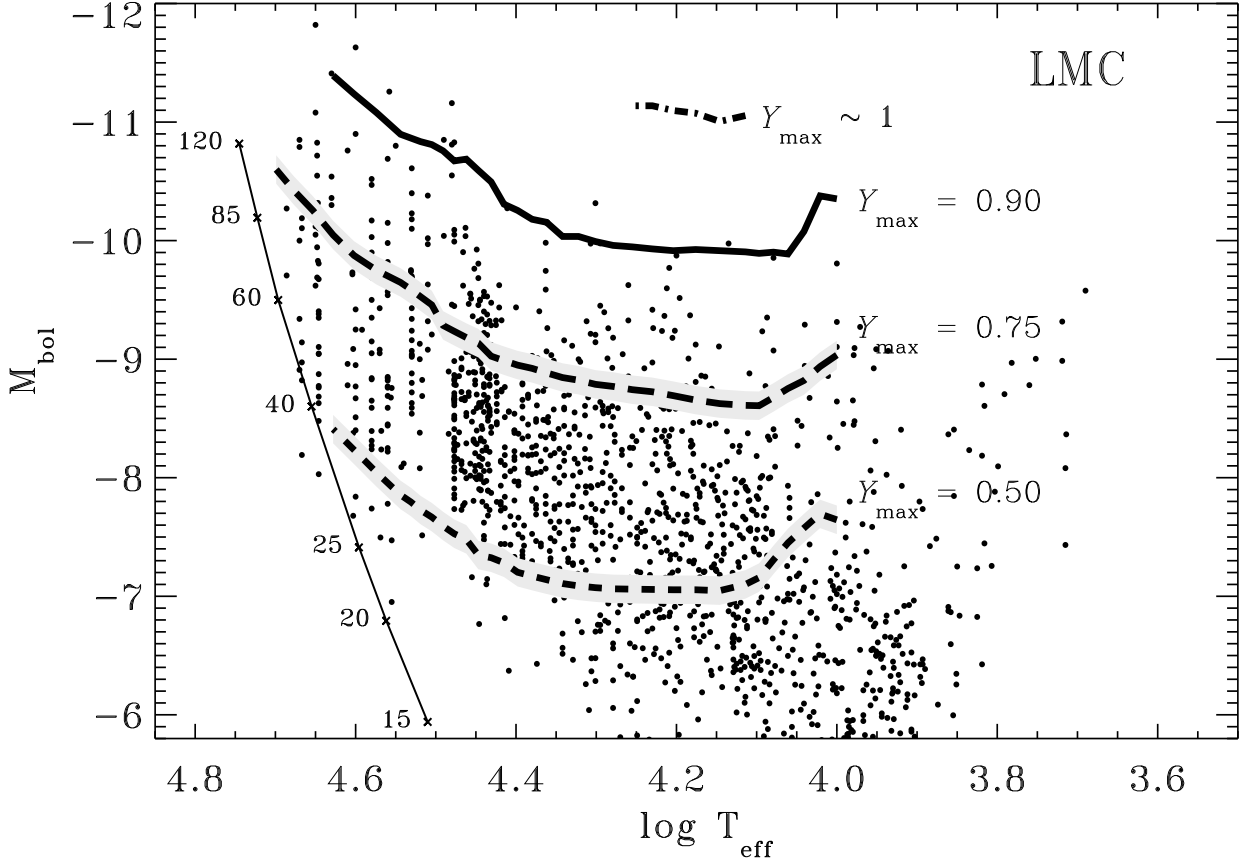}{5.0 in}{90.}{88.0}{90.0}{352.}{-80.}
\caption{\label{lmcfig}
LMC Upper HR diagram from Fitzpatrick and Garmany (1990) with additional
O stars from Walborn and Blades (1997).
The thick solid and dashed lines show the locus
of atmosphere models for
$Y_{\rm max} = g_{\rm rad,max}/g_{\rm grav}$ = 1.0, 0.90,
0.75, and 0.50.
The locus for $Y_{\rm max}= 0.90$ most closely resembles the upper
luminosity limit.
The points for $Y_{\rm max} =1$ were estimated as described in the text,
and show that the modified Eddington limit,
i.e. $Y_{\rm max} = 1$, is about one magnitude higher than the
brightest stars with $T_{\rm eff} \lesssim 25,000$K.
A few stars are above the 0.90~limit, in accord with expectations
of a few misidentified effective temperatures, unresolved binaries,
observational error.
Additionally, the physical depth of the LMC induces
scatter of up to 0.5~mag (assuming the LMC's depth is
comparable to its 10kpc width).
}
\end{figure}

Studies of the upper HR Diagram have often focused on the LMC for
well-known observational reasons, including the uniform and
well-determined distance of the stars, the low line-of-sight reddening,
and the nearly complete census of the most luminous stars.
In Figure~\ref{lmcfig},
we reproduce the LMC HR Diagram published by Fitzpatrick
\& Garmany (1990; small filled circles).  Two changes have been made
for this paper.  First we adjusted the values of $M_{\rm bol}$ to
reflect the currently favored LMC distance modulus of 18.6 mag (e.g.,
Whitelock, van~Leeuwen, \& Feast 1997).  Second, we added data for
about 80 O stars near the 30 Doradus region from a recent paper by
Walborn and Blades (1997).  The various
features of the LMC HR Diagram, and the details of its construction,
are discussed by Fitzpatrick \& Garmany.  For our purposes here, the
important aspect is that there are many stars more luminous than
$M_{\rm bol} = -10 {\rm ~for~} T_{eff} \gtrsim 25,000$~K
while there are few,
if any, for $T_{eff} \lesssim 25,000$~K (including the M supergiants,
which are not shown), i.e., the temperature-dependent HD-limit.
  
In Figure 3 we also show the results of the stellar atmosphere
calculations for $Y_{\rm max}$ = 1, 0.90, 0.75, and 0.50.
These were computed for $Z/Z_\odot = 0.3$,
appropriate for the LMC, and converted to the HR Diagram using a grid
of stellar evolution models interpolated between the Schaerer et al.
(1993a, $Z/Z_{\odot}$ = 0.4) and Charbonnel et al.  (1993; $Z/Z_{\odot}$
= 0.2) grids.  The luminosity at $Y_{\rm max} = 1$ was estimated only
for temperatures in the range 13,000--18,000 K, for which stable models
could be computed out to $Y_{\rm max} \simeq 0.98$.  The extrapolation
to the modified Eddington limit ($Y_{\rm max} = 1$) is thus relatively
secure in this region.  The temperature dependence of the modified
Eddington limit outside these temperatures may be inferred from the
shapes of the other curves.

Figure 3 shows that the bottom of the trough of the modified Eddington
limit ($M_{\rm bol} \simeq$ $-11$) is about one magnitude more luminous
than the brightest LMC stars with $T_{eff} \lesssim 25,000$K.  This
result is actually quite similar to those found by LF and LN; however
in those papers, known deficiencies in the model atmosphere opacities
(LF) and inadequate transformations to the HR Diagram (LF and LN)
obscured the significance of the discrepancy.  Thus, in contrast to
previous assertions, we believe that Figure 3 shows quite clearly that
{\it the modified Eddington limit does not coincide with the
observed upper luminosity limit for LMC stars.} Rather, we suggest the
observed limit is much better defined by the model atmospheres with
$Y_{\rm max} = 0.90$, with their lowest luminosities
at $M_{\rm bol} =$ --9.9.

We can estimate the metallicity dependence of the upper luminosity
limit, as in LN, by comparing the luminosities of the $Y_{\rm max}$ =
0.90 models at various metallicities.  Figure 4 shows such a comparison
for $Z/Z_{\odot}$ = 2.0, 1.0, 0.3, and 0.1, corresponding approximately
to M31, the Milky Way, LMC, and SMC, respectively.  The curves for the
four metallicities have nearly identical shapes, ``the Eddington
trough,'' and differ only by simple displacements.  From this
comparison we might expect the luminosity limit for cool stars in the
SMC to be higher by 0.3 mag than the LMC, and those for the Milky Way
and M31 to be lower by 0.4 and 0.75 mag, respectively.  The relatively
small expected difference between the LMC and SMC is consistent with
the lack of any obvious offset between the LMC and SMC HR Diagrams
(e.g., Garmany \& Fitzpatrick 1989).
Despite the observational challenge in resolving individual stars in M31
(cf. Massey~et~al.~1995), 
comparison of that galaxy with the LMC or SMC offers the best hope for testing
the predictive value of the $Y_{\rm max} = 0.9$ curves.
Additionally, a strong metallicity gradient exists in M31
(a factor of $\sim 5$ from the center to 20 kpc; Blair, Kirshner,
\& Chevalier 1981, 1982), so it may be possible to observe
the variation of the upper-luminosity within the galaxy.
For these purposes, the relevant indicator of the
metallicity as discussed in this paper would be the
Fe/H ratio because, we believe,
the important variation in opacity with metallicity
is due primarily to iron peak elements which blanket the UV.

\section{Concluding Comments \label{concomsec}}

In summary, we have determined the location of the modified Eddington
limit for stars in the LMC using the most recent atmosphere models
combined with a precise mapping to the HR Diagram through up-to-date
stellar evolution calculations.  We find that the modified Eddington
limit is actually {\it a full magnitude higher} than the upper
luminosity limit observed for LMC stars.  The observed limit is
consistent with atmosphere models in which the maximum value of the
ratio of the radiation force outwards to the gravitational force
inwards, $Y_{\rm max}$, is 0.9; i.e., the photospheres of stars at
the observed luminosity limit are bound.  

With some caution, we thus suggest that the simple picture in which a
massive star evolves redward until its photosphere reaches the modified
Eddington limit and becomes unbound is invalid.  Although the stars do
evolve from the Zero Age Main Sequence in the direction of increasing
$Y_{\rm max}$, an instability evidently sets in {\it before} the
atmospheres reach the formal modified Eddington limit at $Y_{\rm max} =
1.0$.  This conclusion is necessarily tentative since this analysis,
like others before, relies on plane-parallel, hydrostatic atmosphere
models, while the atmospheres of real stars near the observed
luminosity limit are likely to share neither of these properties.  It
appears unlikely, however, to be a coincidence that the temperature
dependence of the luminosity limit should so closely match that of the
$Y_{\rm max}$ curves seen in Figures 1--4, whose shapes are nearly
invariant to metallicity or to the precise value of $Y_{\rm max}$ itself.
The degree of stability against radiation pressure of the photospheres
clearly plays an important role in shaping the upper stellar luminosity
limits, although the current characterization of that stability may
leave something to be desired.  The $Y_{\rm max}$ parameterization may
well turn out to correlate with some more critical property, such as
the depth of the ``boundary'' between a stellar wind and the underlying
photosphere.  

A firm understanding of the upper luminosity limits and of the
outbursts in LBVs will almost certainly require a melding of stellar
wind, stellar photosphere, and stellar evolution calculations.
Fortunately, progress in this area is being made (e.g., \cite{sel93},
Schaerer~et~al.~1996).

\acknowledgements

We thank Bohdan Paczy\'nski for helpful comments.
AU was supported by an NSF graduate fellowship and NSF
grants AST93-13620 and AST95-30478.

\newpage

\begin{figure}
\plotfiddle{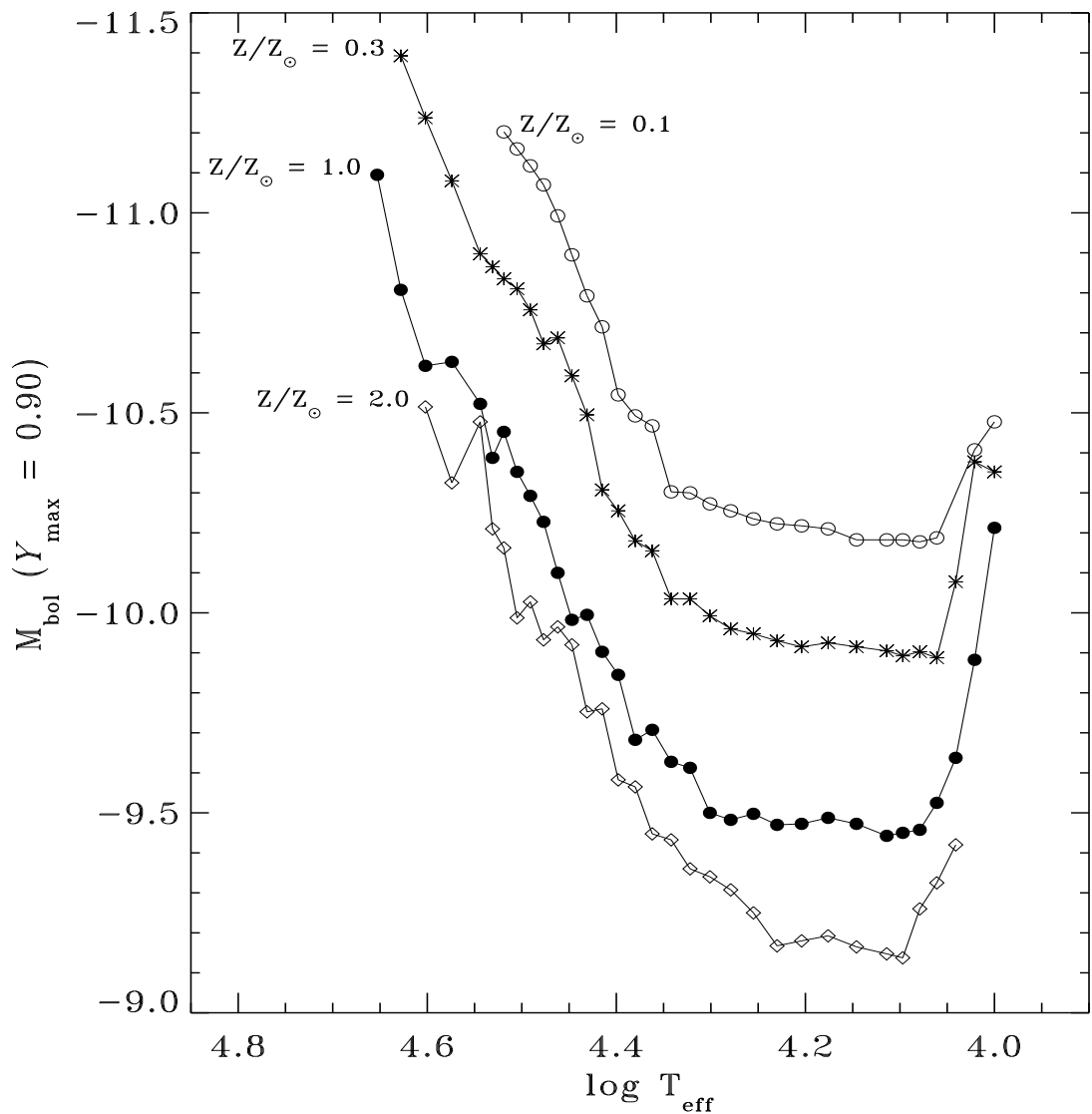}{5.2 in}{0.}{85.0}{75.0}{-270.0}{-100.0}
\caption{\label{metfig}
The metallicity dependence of the luminosity limit corresponding to
$Y_{\rm max}=0.90$ for 
($\log(Z/Z_{\odot}) =$~1.3, 1, -0.5, and -1,
which are appropriate for M31, our galaxy, the LMC, and the SMC,
respectively. The shape of the curves, ``the Eddington trough,''
remains nearly constant as a function of metallicity.}
\end{figure}

\end{document}